\documentclass[12pt,preprint]{aastex}

\def\Gb{\Gamma}
\def\bb{\beta}
\def\Gej{\Gamma_{\rm ej}}
\def\rhoej{\rho_{\rm ej}}
\def\bej{\beta_{\rm ej}}
\def\pf{p_f}
\def\pr{p_r}
\def\ur{u_r}
\def\hf{h_f}
\def\hr{h_r}
\def\rf{r_f}
\def\rr{r_r}

\def\betf{\beta_f}
\def\br{\beta_r}
\def\rhof{\rho_f}
\def\rhor{\rho_r}
\def\g43{\gamma_{43}}
\def\b43{\beta_{43}}
\def\tobs{t_{\rm obs}}
\def\tpl{t_{\rm plateau}}

\def\be{\begin{equation}}
\def\ee{\end{equation}}
\def\beq{\begin{eqnarray}}
\def\eeq{\end{eqnarray}}

\newbox\grsign \setbox\grsign=\hbox{$>$} \newdimen\grdimen \grdimen=\ht\grsign
\newbox\simlessbox \newbox\simgreatbox \newbox\simpropbox
\setbox\simgreatbox=\hbox{\raise.5ex\hbox{$>$}\llap
     {\lower.5ex\hbox{$\sim$}}}\ht1=\grdimen\dp1=0pt
\setbox\simlessbox=\hbox{\raise.5ex\hbox{$<$}\llap
     {\lower.5ex\hbox{$\sim$}}}\ht2=\grdimen\dp2=0pt
\setbox\simpropbox=\hbox{\raise.5ex\hbox{$\propto$}\llap
     {\lower.5ex\hbox{$\sim$}}}\ht2=\grdimen\dp2=0pt

\begin{document}

\title{Mechanical Model for Relativistic Blast Waves}

\author{Andrei M. Beloborodov,\altaffilmark{1} Zuhngwhi Lucas Uhm}
\affil{Physics Department and Columbia Astrophysics Laboratory, \\
       Columbia University, 538 West 120th Street, New York, NY 10027.}

\altaffiltext{1}{Also at Astro-Space Center, Lebedev Physical Institute, 
Profsojuznaja 84/32, Moscow 117810, Russia}

\begin{abstract}
Relativistic blast waves can be described by a mechanical model. 
In this model, the ``blast'' --- the compressed gas between the forward and 
reverse shocks --- is viewed as one hot body. Equations governing its 
dynamics are derived from conservation of mass, energy, and momentum. 
Simple analytical solutions are obtained in the two limiting cases of 
ultra-relativistic and non-relativistic reverse shock.
Equations are derived for the general explosion problem.
\end{abstract}



\section{Introduction}

Relativistic blast waves are believed to produce afterglow emission of 
gamma-ray bursts (GRBs). The blast wave is driven by a shell with Lorentz 
factor $\Gej\sim 10^2-10^3$ which is ejected by a central trigger of the 
explosion. The ejecta expands and drives a forward shock (FS) into the 
external medium, and a reverse shock (RS) propagates inside the ejecta.
This standard explosion picture has four regions: 1 --- external medium, 
2 --- shocked external medium, 3 --- shocked ejecta, and 4 --- unshocked 
ejecta (e.g. Piran 2004). 

The full hydrodynamical simulations of ultra-relativistic blast waves are 
expensive, while GRB modeling requires exploration of a broad range of 
models. Therefore, approximate hydrodynamical calculations are commonly used.
A customary approximation divided the explosion into two stages: before and 
after the RS crosses the ejecta.
At first stage, pressure balance across the blast wave was assumed, i.e. 
the pressures at FS and RS were equated, $\pf=\pr$.
At the second stage, the self-similar solution of Blandford \& McKee 
(1976) was applied. This description, however, has two drawbacks: 
(1) the approximation $\pf=\pr$ violates energy conservation for 
adiabatic blast waves (by as much as a factor of 3 in some example
explosions that we studied), and (2) the model has to assume a sharp
trailing edge of the ejecta, so that $\pr$ suddenly drops to zero when 
RS reaches the edge. 
A realistic model should describe the transition to regime $\pr\ll \pf$;
the transition is especially interesting if the ejecta has a long tail.

In this paper, we propose a ``mechanical'' model which does not have these 
problems. Like the previously used approximation, we assume that regions 2 
and 3 move with a common Lorentz factor $\Gb$.
Thus, in essence we replace the gas between 
FS and RS by one body, which we call ``blast.'' This is reasonable
since internal motions in the blast are subsonic, 
and hydrodynamical simulations confirm that $\Gb\approx const$ between the 
FS and RS (Kobayashi \& Sari 2000).
Instead of finding $\Gb$ from the condition $\pf=\pr$, we leave $\Gb$ as a 
dynamical variable, and find its evolution from a differential equation 
expressing momentum conservation. For a given instantaneous 
$\Gb$, pressures $\pf$ and $\pr$ are found from the jump conditions at the 
two shock fronts, and the difference $(\pf-\pr)$ governs the evolution
of $\Gb$.

Our model assumes spherical symmetry. However, it remains valid if the 
explosion is driven by a jet with a small opening angle $\theta_{\rm jet}$
as long as $\Gb\gg\theta_{\rm jet}^{-1}$. Then the jet behaves like a 
portion of spherical ejecta since the edge of the jet is causally 
disconnected from its axis.

In \S~2, we present a useful form of the jump conditions for a shock with 
arbitrary Lorentz factor and apply them to the FS and RS of a relativistic 
blast wave. The equations of mechanical model and analytical solutions
are derived in \S~3.


\section{JUMP CONDITIONS}

Ideal gas has 
stress-energy tensor $T^{\alpha\beta}=(e+p)u^\alpha u^\beta+g^{\alpha\beta}p$
and mass flux $j^\alpha=\rho u^\alpha$, where
$g^{\alpha\beta}$ is Minkowski metric,
$u^\alpha$ is 4-velocity of the gas, $\rho$ is rest-mass density, 
$e$ is energy density (including rest energy), and $p$ is pressure; all
the thermodynamic quantities are defined in the rest frame of the gas. 
A shock is described by three jump conditions that express 
continuity of $j^\alpha$ and $T^{\alpha\beta}$ in the shock frame
(Landau \& Lifshitz 1959),
$$
\gamma_2 \beta_2\, \rho_2 = \gamma_1 \beta_1\, \rho_1, \qquad
\gamma_2^2 \beta_2\, (e_2+p_2) = \gamma_1^2 \beta_1\, \rho_1 c^2, \qquad
\gamma_2^2 \beta_2^2\, (e_2+p_2) + p_2 = \gamma_1^2 \beta_1^2\, \rho_1 c^2.
$$
Here subscripts 1 and 2 refer to 
preshock (cold) and postshock (hot) medium, $\beta_1$ and $\beta_2$ are the 
gas velocities relative to the shock front, and we assumed $p_1=0$ and 
$e_1=\rho_1c^2$. 
The shock strength may be described by relative velocity 
$\beta_{12}=(\beta_1-\beta_2)/(1-\beta_1\beta_2)$ 
or $\gamma_{12}=(1-\beta_{12}^2)^{-1/2}$. 
The postshock gas satisfies $e_2/\rho_2c^2=\gamma_{12}$, and a 
convenient approximation for $p_2$ is
\be
\label{eq:kap}
  p_2=\frac{1}{3}\left(1+\frac{1}{\gamma_{12}}\right)(e_2-\rho_2c^2).
\ee
It is exact if the gas is monoenergetic, i.e., 
particles have equal energies in the gas frame. Its error for Maxwellian gas 
is within 5\%. Using equation~(\ref{eq:kap}), we express all quantities 
in terms of $\beta_{12}$ (or $\gamma_{12}$), which remains as the only free 
parameter of the shock,
\be
\label{eq:jump}
 \beta_2=\frac{\beta_{12}}{3}, \qquad \rho_2 = 4 \gamma_{12}\rho_1,
 \qquad e_2 = 4 \gamma_{12}^2\, \rho_1 c^2, 
 \qquad p_2 = \frac{4}{3}\, (\gamma_{12}^2-1)\, \rho_1 c^2.
\ee
These equations apply to shocks of arbitrary strength, relativistic or 
non-relativistic.

The blast wave has two shocks, forward and reverse. 
The above equations with $\gamma_{12}=\Gb$ describe FS. The RS is described 
by the same equations when index 1 is replaced by 4 and index 2 is replaced 
by 3. Pressures $\pf=p_2$ and $\pr=p_3$ are given by
\be
\label{eq:p}
  \pf= \frac{4}{3}\,(\Gb^2-1)\, \rho_1 c^2, \qquad
  \pr=\frac{4}{3}\, (\gamma_{43}^2-1)\, \rho_4 c^2.
\ee 
The approximation of common Lorentz factor of regions 2 and 3 implies that 
$\Gb$ and $\g43$ are not independent. We assume that the Lorentz factor of 
unshocked ejecta, $\Gej$, is known in the lab frame. 
Then one finds its Lorentz factor relative to the blast, 
$\gamma_{43}=\Gb\Gej(1-\bb\bej)$.  For relativistic blast 
waves ($\Gej\gg 1$ and $\Gb\gg 1$) this relation simplifies,
\be 
\label{eq:g43}
  \gamma_{43}=\frac{1}{2} \left(\frac{\Gej}{\Gb}+\frac{\Gb}{\Gej}\right),
  \qquad \b43=\frac{\Gej^2-\Gb^2}{\Gej^2+\Gb^2}.
\ee
Thus, given the parameters of regions 1 and 4, only one parameter of the 
blast wave is left free --- $\Gb$.
In particular, denoting $\rhoej=\rho_4$, we find
\begin{equation}
  \frac{\pr}{\pf}=\frac{\rhoej}{4\Gej^2\rho_1}
                  \left(\frac{\Gej^2}{\Gb^2}-1\right)^2.
\end{equation}
If the pressure balance $\pf=\pr$ is assumed, it immediately determines 
the instantaneous $\Gb$,
\be 
 \Gb=\Gej\left[1+2\Gej\left(\frac{\rho_1}{\rhoej}\right)^{1/2}\right]^{-1/2}, 
  \qquad \pf=\pr.
\ee
The problems of this approximation have been mentioned in \S~1. In particular, 
energy conservation is not satisfied and requires a different solution.


\section{MECHANICAL MODEL}

The gas in the blast wave flows radially with 4-velocity
$u^\alpha=(\gamma,\gamma\beta,0,0)$ in spherical coordinates 
$(t,r,\theta,\phi)$. Rest-mass conservation 
$\nabla_\mu(\rho u^\mu)=r^{-2}\partial_\mu (r^2\rho u^\mu)=0$ gives
\be
\label{eq:rho}
  \frac{1}{r^2}\frac{d}{dt}\left(r^2\rho\gamma\right)
   +\rho\gamma\,\frac{\partial \beta}{\partial r}=0,
\ee
where $\frac{d}{dt}\equiv\frac{\partial }{\partial t} 
+\beta\frac{\partial }{\partial r}$
is the convective derivative (hereafter we use units $c=1$).
Four-momentum conservation $\nabla_\mu T_\alpha^{\;\mu}=0$ gives two 
independent equations ($\alpha=0,1$)
\be
\label{eq:ua}
  \frac{1}{r^2}\frac{d}{dt}\left(r^2h\gamma u_\alpha\right) 
   +h\gamma u_\alpha\frac{\partial\beta}{\partial r}+\partial_\alpha p=0,
\ee
where $h=e+p$. 
Instead of $\nabla_\mu T_0^{\;\mu}=0$ we will use the projection
$u^\alpha \nabla_\mu T_{\alpha}^{\;\mu}=0$ which yields
$r^{-2}\partial_\mu(r^2 hu^\mu)=\gamma dp/dt$.
So, conservation of 4-momentum is expressed by
\beq
\label{eq:cons}
   \frac{1}{r^2}\frac{d}{dt}\left(r^2h\gamma^2\beta\right)
          = -\frac{\partial p}{\partial r}
           -h\gamma^2\beta\frac{\partial\beta}{\partial r}, \qquad
   \frac{1}{r^2}\frac{d}{dt}\left(r^2h\gamma\right)
  = \gamma\,\frac{d p}{d t}-h\gamma\,\frac{\partial\beta}{\partial r}.
\eeq
We apply equations~(\ref{eq:rho}) and (\ref{eq:cons}) to 
the gas between FS and RS and make the approximation 
\be
  \gamma(t,r)=\Gb(t), \qquad \partial\beta/\partial r=0, \qquad \rr<r<\rf,
\ee
where $\rr(t)$ and $\rf(t)$ are the instantaneous radii of RS and FS, 
respectively. Then the
integration of equations~(\ref{eq:rho}) and (\ref{eq:cons})
over $r$ between RS and FS (at $t=const$) yields
\beq
\label{eq:1}
   \frac{\Gb}{r^2}\frac{d}{dr}\left(r^2\;\Sigma\;\Gb\right) & = & 
  \rhor(\bb-\br)\Gb^2+\frac{1}{4}\,\rhof, \\
\label{eq:2}
   \frac{1}{r^2}\frac{d}{dr}\left(r^2H\Gb^2\right)
    & = & \hr\,(\bb-\br)\Gb^2+\pr,\\
\label{eq:3}
   \frac{\Gb}{r^2}\frac{d}{dr}\left(r^2H\;\Gb\right)
        & = & \Gb^2\frac{dP}{dr}+(\hr-\pr)(\bb-\br)\Gb^2+\frac{3}{4}\,\pf,
\eeq
where $\Sigma\equiv\int_{\rr}^{\rf}\rho\,dr$, $H\equiv\int_{\rr}^{\rf} h\,dr$,
$P\equiv\int_{\rr}^{\rf} p\,dr$, $\br=d\rr/dt$, and $\betf=d\rf/dt$. 
In the derivation of equations~(\ref{eq:1})-(\ref{eq:3}) we made use of
$\Gamma\gg 1$: (1)
the relativistic blast is a very thin shell, $\rf-\rr\sim r/\Gb^2\ll r$,
so we used $\rf\approx\rr\approx r$ when calculating the integrals, 
(2) we used the jump condition at the FS
$\betf-\bb=1/4\Gb^2$ and $\hf=4\pf\gg \rhof$, 
(3) the convective derivative $d/dt$ has been replaced by 
$\bb d/dr\approx d/dr$ and $\Gb^2\bb$ by $\Gb^2$ in equation~(\ref{eq:2}).
Besides, the last term $\rho_f/4$ in equation~(\ref{eq:1}) may be neglected.

RS may not be relativistic.
We derive from the jump conditions for a RS with $\g43\ll\Gamma$, 
\be
\label{eq:br}
   \bb-\br=\frac{\Gej^2-\Gb^2}{2\Gb^2(\Gej^2+2\Gb^2)},
\ee
\be
\label{eq:RS}
   \rhor=2\left(\frac{\Gej}{\Gb}+\frac{\Gb}{\Gej}\right)\rhoej,  \qquad
   \pr=\frac{1}{3}\left(\frac{\Gej}{\Gb}-\frac{\Gb}{\Gej}\right)^2\rhoej,\qquad
\hr=\frac{4}{3}\left(\frac{\Gej^2}{\Gamma^2}+\frac{\Gamma^2}{\Gej^2}+1\right)
     \rhoej.
\ee
This leaves four unknowns in equations~(\ref{eq:1})-(\ref{eq:3}): 
$\Sigma$, $H$, $P$, and $\Gb$. One more equation is required to close
the set of equations. We propose the following approximate relation,
\be
\label{eq:4P}
   H-\Sigma=4P.
\ee  
As shown below, it is accurate in both limits of ultra-relativistic 
and non-relativistic RS, and should be a reasonable approximation in 
an intermediate case.

\subsection{Relativistic Reverse Shock}

In the case of relativistic RS, $\g43\gg 1$, the gas is relativistically 
hot in both regions 2 and 3, and obeys the equation of state $h=4p$ 
everywhere in the blast. Then $\Sigma$ may be neglected in 
equation~(\ref{eq:4P}) and the condition $H=4P$ closes the set of 
mechanical equations.
Besides, we have $\bb-\br=(2\Gb^2)^{-1}$ (since $\Gej\gg\Gb$, see 
eq.~\ref{eq:br}). Equations~(\ref{eq:1})-(\ref{eq:3}) then read
$$
    \frac{\Gb}{r^2}\frac{d}{dr}\left(r^2\Sigma\Gb\right) 
    = \frac{1}{2}\,\rhor+\frac{1}{4}\,\rhof,  \quad
    \frac{1}{r^2}\frac{d}{dr}\left(r^2H\Gb^2\right) = 3\pr, \quad
    \frac{\Gb}{r^2}\frac{d}{dr}\left(r^2H\Gb\right)
    = \frac{\Gb^2}{4}\,\frac{dH}{dr}+\frac{3}{2}\,\pr+\frac{3}{4}\,\pf.
$$
From the last two equations we find $H(r)$, then substitute to the 
second equation and get the final differential equation for $\Gb$, 
\be
   \frac{dH}{dr}+\frac{4H}{r}=4\rho_1, \qquad
  H(r)=\frac{4}{r^4}\int_0^r\rho_1 r^4 dr.
\ee
\be
   \frac{r}{\Gb}\,\frac{d\Gb}{dr}
   =\frac{r^5}{8}\left(\int_0^r\rho_1 r^4 dr\right)^{-1}
    \left(\frac{\Gej^2\rhoej}{\Gb^4}-4\rho_1\right)+1.
\ee
We have used here the jump condition $\pr=\frac{1}{3}\,(\Gej/\Gb)^2\rhoej$ 
for the relativistic RS (see eq.~\ref{eq:RS}).
For external medium with a power-law density profile $\rho_1\propto r^{-k}$, 
one can simplify $\int\rho_1 r^4 dr=r^5\rho_1/(5-k)$.
For example, if $\rho_1\propto r^{-2}$ (wind-type external medium), 
$\Gej=const$ and $\rhoej\propto r^{-2}$, then the solution is $\Gb=const$ 
and we find
\be
   \pf=3\pr, \qquad \Gb=\left(\frac{3\Gej^2\rhoej}{4\rho_1}\right)^{1/4}, 
   \qquad k=2.
\ee

\subsection{Non-relativistic Reverse Shock}

In the case of non-relativistic RS, $\b43\ll 1$, we have 
$$
  \rhor=4\rhoej, \qquad \pr=\frac{4}{3}\,\b43^2\rhoej, \qquad 
  \b43=1-\frac{\Gb}{\Gej}, \qquad \beta-\br=\frac{\b43}{3\Gej^2},
$$
and the 
equation of state $\ur=\frac{3}{2}\pr$ and $\hr=\rhor+\ur+\pr\approx\rhor$.
The effective inertial mass of the blast $H$ is dominated by region~3 
as long as $\pr/\pf\gg\b43$. Indeed, the thicknesses of regions~2 and 3 
are $\sim r/\Gb^2$ and $\sim\b43 r/\Gb^2$, respectively, and the 
contributions to $H$ from regions 2 and 3 are 
$H_2\sim\hf r/\Gb^2$ and $H_3\sim \rhor\b43 r/\Gb^2$. This gives
$H_2/H_3\sim(\pf/\pr)\b43\ll 1$. 
On the other hand, $H-\Sigma$ and integrated pressure $P$ of the blast is
dominated by region 2. In particular, $P_2/P_3\sim(\pf/\pr)\b43^{-1}\gg 1$.
Therefore the relation $H-\Sigma=4P$ 
remains correct for blast waves with non-relativistic RS.
Using this relation and
subtracting equation~(\ref{eq:1}) from (\ref{eq:3}), we derive the 
equation for $P$ in the leading order of $\b43$,
\be
\label{eq:Pnr}
 \frac{4}{r^2\Gej}\frac{d}{dr}\left(r^2P\Gej\right)=\frac{dP}{dr}+\rho_1.
\ee
Parameters of the RS do not enter this equation.
The ejecta plays the role of a piston that sweeps up
the external medium with $\Gb\approx\Gej$, and equation~(\ref{eq:Pnr})
in essence describes the shocked external medium ahead of the piston.
In the case of $\Gej=const$, equation~(\ref{eq:Pnr}) has the solution,
\be
\label{eq:P}
   P(r)=\frac{1}{3}\,r^{-8/3}\int_0^r r^{8/3}\rho_1(r)\,dr.
\ee
 
Subtracting equation~(\ref{eq:3}) from (\ref{eq:2}), we find
$H\Gb\frac{d\Gb}{dr}=-\Gb^2\frac{dP}{dr}+\pr(\bb-\br)\Gb^2+\pr
  -\frac{3}{4}\pf$, which gives, in the leading order of $\b43$,
\be
\label{eq:dbet}
  \Sigma\frac{d\b43}{dr}=-\frac{8}{3}\,\frac{P}{r}+\frac{4}{3}\,\rho_1
                         -\frac{\pr}{\Gej^2}.
\ee
Finally, equation~(\ref{eq:1}) in the leading order becomes
\be
\label{eq:dSig}
  \frac{1}{r^2}\,\frac{d}{dr}\left(r^2\Sigma\right)
   =\frac{4}{3}\,\rhoej\frac{\b43}{\Gej^2}.
\ee
From equations~(\ref{eq:dbet}) and (\ref{eq:dSig}) we find
\be
\label{eq:betSig}
   \frac{1}{r^2}\frac{d}{dr}\left(r^2\Sigma\,\b43\right)=-\frac{8}{3}\,
         \frac{P}{r}+\frac{4}{3}\,\rho_1.
\ee

Analytical solutions can be found for $\b43$ and $\Sigma$ when
the external medium has a power-law density profile, $\rho_1\propto r^{-k}$,
$\Gej=const$, and $\rhoej\propto r^{-2}$.
Then $P(r)=\rho_1r/(11-3k)$, 
\be
   \Sigma^2=\frac{32}{3}\,\frac{\rho_1\rhoej r^2}
             {(11-3k)(4-k)\Gej^2}, \qquad
    \b43^2=\frac{3}{2}\frac{(4-k)\Gej^2\rho_1}
          {(11-3k)\rhoej},
  \qquad
     \frac{\pf}{\pr}=\frac{2}{3}\left(\frac{11-3k}{4-k}\right).
\ee


\section{DISCUSSION}

The RS may change from non-relativistic to relativistic as the blast wave 
propagates. The mechanical model formulated in 
equations~(\ref{eq:1})-(\ref{eq:4P}) is applicable to such a general case. 
The solution can be found numerically for any given 
external medium $\rho_1(r)$ and ejecta $\Gej(t,r)$ and $\rhoej(t,r)$.
Explosions in media with inhomogeneities can be studied, 
as well as ``refreshed'' explosions with an energetic tail of the ejecta. 
Examples of such blast waves and their afterglow emission will be presented 
elsewhere (Uhm \& Beloborodov, in preparation).

The RS shock becomes unimportant for the blast-wave dynamics when 
$\pr\ll\pf$. This transition is consistently described by the mechanical 
model. The time of the transition depends on the structure of 
ejecta. Recent {\it Swift} observations of the plateau in the X-ray light 
curve at $\tobs<\tpl\sim 10^3-10^4$~s suggest that 
an energetic tail of the ejecta keeps pushing the blast as long as 
$10^4$~s in observer's time $\tobs\sim r/2c\Gb^2$ (Nousek et al. 2006).

The solution at late stages with $\pr\ll\pf$ is obtained from dynamical 
equations~(\ref{eq:2})-(\ref{eq:3}) by setting $p_r=0$. In particular, for 
external medium with $\rho_1\propto r^{-k}$, we find
\be
    \Gb^2=\frac{(5-k)E}{16\pi\rho_1r^3}, \qquad \pr\ll\pf,
\ee
where $E$ is the isotropic energy of explosion. This 
is close to the exact self-similar solution of the full
hydrodynamical equations (Blandford \& McKee 1976) except the 
numerical factor: 
$\Gb$ of the mechanical model is lower than the exact Lorentz factor
behind FS by a factor of $[(5-k)/(17-4k)]^{1/2}$. This difference 
is caused by the steep profile of the gas Lorentz factor $\gamma(t,r)$ 
behind the FS at late stages. Then the approximation of mechanical model 
($\partial\gamma/\partial r=0$ between RS and FS)
becomes less accurate, however, still gives a reasonable solution.
At later stages, the gas motion becomes non-radial and the blast-wave 
equations must include the $\theta$-dependence (e.g. Kumar \& Granot 2003).

The mechanical model
admits a similar formulation for the case of strongly magnetized
ejecta (Zhang \& Kobayashi 2005 and refs. therein) if the jump conditions
are changed accordingly.
It can be extended to include the neutron component of the ejecta
(Derishev, Kocharovsky, \& Kocharovsky 1999; Beloborodov 2003 and 
refs. therein) and the effects of $e^\pm$-loading and preacceleration
of external medium by the prompt $\gamma$-rays 
(Thompson \& Madau 2000; Beloborodov 2002). 

\acknowledgements
This work was supported by NASA grant NAG5-13382 and Swift grant. 



\end{document}